\documentclass[10pt]{iopart}

\usepackage{graphicx}
\usepackage{iopams}
\usepackage{textcomp}
\usepackage{citesort}

\newcounter{parentequation}

\newenvironment{subequations}{%
  \refstepcounter{equation}%
  \setcounter{parentequation}{\value{equation}}%
  \setcounter{equation}{0}%
  \def\theequation{\theparentequation\alph{equation}}%
  \ignorespaces
}{%
  \setcounter{equation}{\value{parentequation}}%
  \ignorespacesafterend
}

\begin{document}

\title[Dynamics of paramagnetic superrotors in external magnetic fields]{A theoretical study of the dynamics of paramagnetic superrotors in external magnetic fields}

\author{Johannes Flo\ss}
\address{Department of Chemical Physics, Weizmann Institute of Science, 234 Herzl Street, Rehovot 76100, Israel}
\ead{johannes.floss@weizmann.ac.il}

\begin{abstract}
We present a detailed theoretical study of oxygen molecules in high rotational states (molecular superrotors) interacting with an external magnetic field.
The system shows rich dynamics, ranging from a spin-selective splitting of the angular distribution over molecular alignment to an inversion of the rotational direction.
We find that the observed magneto-rotational effects are due to a spin-mediated precession of the orbital angular momentum around the magnetic field.
Analytical expressions for the precession frequency in the limits of weak and strong magnetic fields are derived and used to support the proposed mechanism.
In addition, we provide the procedure for a numerical treatment of oxygen superrotors in an external magnetic field.
\end{abstract}



\maketitle
\ioptwocol



\section{\label{sec.intro}Introduction}

The control of molecular rotation by electromagnetic fields has proven to be a powerful tool in a number of rapidly growing fields of molecular science (for a recent review, see~\cite{lemeshko13}).
Aligning the molecular axis or the molecular angular momentum is a key requirement in such diverse fields as attosecond high-harmonic spectroscopy~\cite{kim14,lepine14}, photoelectron spectroscopy~\cite{stolow08}, controlling molecular interactions with atoms~\cite{tilford04}, molecules~\cite{vattuone10}, and surfaces~\cite{kuipers88,zare98}, or altering molecular trajectories~\cite{purcell09,gershnabel10,gershnabel11}.

The alignment of the molecular axis by electromagnetic fields is a well-researched problem~\cite{stapelfeldt03,seideman05,ohshima10,fleischer12}.
Polar and paramagnetic molecules can be aligned by the use of static electric and magnetic fields~\cite{friedrich91,friedrich92}.
Here, the applied field interacts with the permanent dipole moment of the molecule and creates pendular states in which the molecular axis librates~\cite{slenczka94}.
The control can be extended to molecules without a permanent dipole moment by the use of laser pulses~\cite{zon75,friedrich95,friedrich95b,seideman95} with intensities of the order of several TW/cm$^{2}$.
The electric field of the laser pulse polarizes the molecule and subsequently interacts with the induced dipole.
Orientation of the angular momentum, i.e. the creation of an ensemble of unidirectionally rotating molecules, can be achieved by schemes like the molecular propeller~\cite{fleischer09,kitano09} which uses two well-timed cross-polarized laser pulses, chiral pulse trains~\cite{zhdanovich11} consisting of periodically applied laser pulses, or the so-called optical centrifuge~\cite{karczmarek99,villeneuve00}.
In the latter technique, the molecules are subject to two counter-rotating circularly polarized fields that are linearly chirped with respect to each other.
The resulting interaction potential creates an accelerated rotating trap, bringing the molecules to a fast-spinning state.

Very fast spinning molecules, with rotational energies comparable to the molecular bond strength, are known as molecular superrotors.
Such states can be created by the above-mentioned optical centrifuge.
An alternative method is to use a train of short laser pulses that are separated by the rotational period~\cite{cryan09}, using the rotational quantum resonance effect~\cite{izrailev80}.
However, the latter scheme is limited because of the centrifugal distortion of the fast spinning molecules~\cite{floss14}.
Due to their high angular momenta, reaching hundreds of $\hbar$~\cite{korobenko13}, superrotors can be expected to show a qualitative different behaviour than normal molecular rotors.
In pioneering work~\cite{villeneuve00}, it was shown that the centrifugal forces of the superrotors can be strong enough to break molecular bonds.
It was also found~\cite{milner14d,milner14c} that superrotors are very robust against collisions, showing gyroscopic dynamics over a long time~\cite{khodorkovsky15}, followed by an explosion-like equilibration of the internal energy.

In a recent experiment, the dynamics of paramagnetic superrotors in a static magnetic field was investigated~\cite{milner14b}.
It was found that a gas of paramagnetic superrotors can become optically birefringent when a magnetic field is applied.
This effect was called magneto-rotational birefringence.
As an example for a paramagnetic superrotor molecular oxygen O$_2$ was used, which has a non-zero electronic spin of $S=1$ in its electronic ground state $X^3\Sigma_g^-$.
By means of the optical centrifuge technique, the molecules were excited to unidirectional rotation with angular momenta of up to $100\hbar$ -- ten times higher than for a typical oxygen molecule at room temperature.
A magnetic field of 3~T, applied at 90~degrees to the laser propagation axis (i.e., the axis of rotation), lead to optical birefringence of the molecular gas, indicating alignment of the molecules.
Noteworthy, the considered magnetic field strength was by far too low to induce the observed alignment in a thermal sample of oxygen molecules.
As a mechanism, it was proposed that the magnetic field leads to a precession of the large orbital angular momentum $\bi{N}$ of the centrifuged molecules around the magnetic field.
This precession then leads to the alignment of the molecular axis along the magnetic field.
It was concluded that the interaction of $\bi{N}$ and the magnetic field is mediated via the electronic spin.

In this article, a detailed theoretical analysis of the magneto-rotational effect reported in~\cite{milner14b} is provided.
Further insights using an experimental approach are provided in a companion article~\cite{korobenko15a}.

The present article is structured as follows.
In section~\ref{sec.model}, the theoretical model is introduced.
The effective Hamiltonian of the rovibronic ground state is described, as well as the numerical methods to calculate the rotational dynamics in the optical centrifuge and the probe signal.
Also, the method to extract the angular distribution of the molecular axis from the spin-rotational wave function is explained.
In section~\ref{sec.results}, the results of the numerical simulations are presented, in particular the simulated probe signals as well as the angular distributions.
In section~\ref{sec.mechanism}, the proposed mechanism of the magneto-rotational effect -- spin mediated precession of the orbital angular momentum around the magnetic field -- is explained in detail.
An analytical expression for the precession frequency is derived for the limiting cases of weak and strong magnetic fields, and used to test the validity of the proposed mechanism.



\section{\label{sec.model}Model and numerical methods}

\subsection{Considered scenario}

\begin{figure}
\centering
\includegraphics[width=\linewidth]{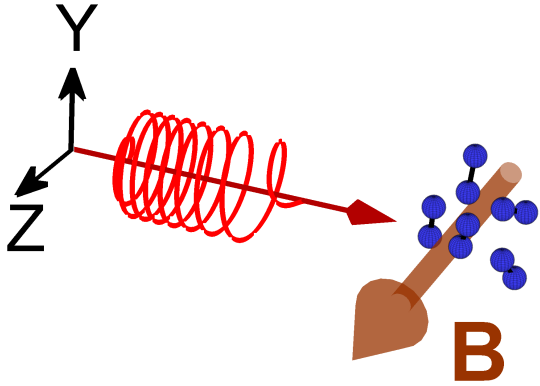}
\caption{
\label{fig.centrifuge}
A sketch of the considered scenario.
An optical centrifuge laser pulse interacts with O$_2$ molecules at room temperature.
Additionally, a static magnetic field $\bi{B}$ directed at 90 degrees to the laser propagation axis is applied.
}
\end{figure}

In this article we consider a scenario resembling the experiment described in~\cite{milner14b}.
It is depicted in figure~\ref{fig.centrifuge}:
An optical centrifuge laser pulse~\cite{karczmarek99,villeneuve00} interacts with molecular oxygen at room temperature.
As a result, the molecules are excited to fast unidirectional rotation in the polarization plane of the laser.
A static magnetic field $\bi{B}$ is applied perpendicularly to the propagation direction of the laser pulse.
The fast rotation of the molecules is measured by a second, circularly polarized laser pulse, which propagates in the same direction as the centrifuge pulse.

In the following, we define the laboratory $z$-axis as the direction of the magnetic field, and the $x$-axis as the laser propagation direction.

In our model, we assume that the intensities of the laser field and the static magnetic field are spatially uniform.
Also, we neglect molecular collisions.
Experimental results have shown that collisional effects and the magneto-rotational birefringence can be well distinguished~\cite{milner14b,milner14c}, justifying this simplification of the model.

\subsection{\label{sec.hamil}Molecular Hamiltonian}

To describe the rotational dynamics of the centrifuged molecules, we use an effective Hamiltonian for the vibronic ground state.
Although the optical centrifuge excites very high rotational states, it was found experimentally~\cite{korobenko13} that for the scenario considered in this work, the rotational levels are still very well described by the Dunham expansion.
We therefore do not need to consider the vibrational states directly, but rather account for the non-rigidity of the molecules by centrifugal distortion terms.

Molecular oxygen has an electronic ground state with two unpaired electrons, and thus a non-zero total electronic spin $\mathbf{S}$; in particular, $S=1$.
This gives rise to spin-spin as well as spin-orbit interactions.

The effective Hamiltonian for the oxygen molecules in the magnetic field is given as~\cite{tinkham55a,tinkham55b,herzberg89,brown03}
\begin{equation}
\mathcal{H}_{\mathrm{eff}}=\mathcal{H}_{\mathrm{rot}}+\mathcal{H}_{\mathrm{ss}}+\mathcal{H}_{\mathrm{so}}+\mathcal{H}_{\mathrm{ms}} \,.
\label{hamil.eff}
\end{equation}
Here, $\mathcal{H}_{\mathrm{rot}}$ describes the rotation of the molecule, $\mathcal{H}_{\mathrm{ss}}$ and $\mathcal{H}_{\mathrm{so}}$ the spin-spin and spin-orbit interactions, respectively, and $\mathcal{H}_{\mathrm{ms}}$ the interaction of the magnetic field with the molecule, in particular with the electronic spin.

The rotational Hamiltonian is given as 
\begin{equation}
\mathcal{H}_{\mathrm{rot}}= B_0 \bi{N}^2 -D \bi{N}^4\,.
\label{hamil.rot}
\end{equation}
The rotational and centrifugal distortion constants are given as $B_0=2.8400\cdot10^{-23}~\mathrm{J}$~\cite{nist} and $D=9.612\cdot10^{-29}~\mathrm{J}$~\cite{nist}, where the former is for the vibronic ground state. 
Note that we use the total orbital angular momentum $\bi{N}$ and not the nuclear orbital angular momentum $\bi{O}$.
They are approximately equal, with small differences arising from the electronic orbital angular momentum~\cite{tinkham55a}.

Since there are unpaired electrons in molecular oxygen, the interaction between the electronic spins does not average to zero.
Tinkham and Strandberg have shown~\cite{tinkham55a} that the two unpaired electrons have a large probability to be found at the same nucleus, and that the spin-spin interaction is minimized when $\bi{S}$ is perpendicular to the molecular axis.
In particular, it is given as
\begin{equation}
\mathcal{H}_{\mathrm{ss}}=\frac{2}{3}\lambda\left[3S_{\zeta}^2-\bi{S}^2\right]\,,
\label{hamil.ss}
\end{equation}
where $S_{\zeta}$ is the projection of the total electronic spin on the intermolecular axis and $\lambda=3.943\cdot10^{-23}~\mathrm{J}$~\cite{brown03} is the spin-spin coupling constant.

The spin-orbit coupling between the electronic spin $\bi{S}$ and the orbital angular momentum $\bi{N}$ is mainly caused by the interaction of the unpaired electrons with the magnetic field of the electrons that follow the rotating nuclei~\cite{tinkham55a}.
It is given as
\begin{equation}
\mathcal{H}_{\mathrm{so}}= \gamma \bi{N}\cdot\bi{S} \label{hamil.sr}\,,
\end{equation}
where $\gamma=-1.674\cdot10^{-25}~\mathrm{J}$~\cite{brown03}.

The by-far strongest interaction between the magnetic field $\bi{B}$ and the molecule is the spin-Zeeman interaction between the field and the electronic spin.
The spin-Zeeman Hamiltonian is given as
\begin{equation}
\mathcal{H}_{\mathrm{ms}}=g_S\mu_B\bi{S}\cdot\bi{B} \,,
\label{eq.spinzeeman}
\end{equation}
with the electronic spin g-factor $g_S=2.002064$~\cite{christensen78} and the Bohr magneton $\mu_B$.
In principle, the magnetic field also interacts directly with the orbital angular momentum $\bi{N}$.
However, the respective g-factor is four orders of magnitude smaller than $g_S$ and therefore negligible on the time-scales considered here.

For the numerical treatment of the problem, it is convenient to work in Hund's case~(b) basis~\cite{townes55,brown03}.
In Hund's coupling case~(b), the electronic spin $\bi{S}$ is coupled to the orbital angular momentum $\bi{N}$ to form the total angular momentum $\bi{J}$.
The basis functions can be denoted as
\begin{equation}
|\phi\rangle=|\eta,\Lambda;N,\Lambda;N,S,J,M\rangle \,.
\end{equation}
Here, $\eta$ is a collective quantum number for the vibronic state, $\Lambda$ is the projection of the electronic orbital angular momentum on the molecular axis, and $M$ is the projection of the total angular momentum on the space-fixed $z$-axis.
Since $\eta$, $\Lambda=0$, and $S=1$ are constant, we will use the short notation $|\eta,\Lambda;N,\Lambda;N,S,J,M\rangle\equiv|JNM\rangle$.
The rotational Hamiltonian is diagonal in Hund's case (b) basis, and the matrix elements are
\begin{equation}
\langle N J M | \mathcal{H}_{\mathrm{rot}} | N J M\rangle = B_0 N(N+1) - D N^2(N+1)^2 \,.
\label{element.rot}
\end{equation}
The non-zero elements of the spin-spin Hamiltonian are
\begin{eqnarray}
\langle N J M |& \mathcal{H}_{\mathrm{ss}} | N' J M\rangle = \nonumber\\
& \lambda\frac{2\sqrt{30}}{3}(-1)^{J+1}\sqrt{(2N+1)(2N'+1)}
\nonumber\\
&\times 
\left\{\begin{array}{ccc}S&N'&J\\N&S&2 \end{array}\right\}
\left(\begin{array}{ccc}N&2&N'\\0&0&0 \end{array}\right) \,,
\end{eqnarray}
where $N'=N,N\pm2$.
The round brackets denote the Wigner 3j-symbol, the curly brackets the Wigner 6j-symbol.
The spin-orbit Hamiltonian is diagonal again, with the elements
\begin{eqnarray}
\langle N J M |& \mathcal{H}_{\mathrm{so}} | N J M\rangle =\nonumber\\
&\sqrt{N(N+1)(2N+1)}
\sqrt{S(S+1)(2S+1)}
\nonumber\\
&\times
\gamma (-1)^{N+J+S}\left\{\begin{array}{ccc}S&N&J\\N&S&1 \end{array}\right\} \,.
\end{eqnarray}
The non-zero matrix elements of the spin-Zeeman term are
\begin{eqnarray}
\langle NJM |& \mathcal{H}_{\mathrm{ms}} | N J' M\rangle=
\nonumber\\
&g_s^e \mu_{\beta} B (-1)^{J+J'-M+S+1+N}
\nonumber\\
&\times\left( \begin{array}{ccc}J & 1 & J'\\-M & 0 & M\end{array}\right)
 \left\{ \begin{array}{ccc}S & J' & N\\J & S & 1\end{array}\right\}
\nonumber\\
&\times\sqrt{(2J'+1)(2J+1)S(S+1)(2S+1)} \,,
\end{eqnarray}
where $J'=J,J\pm1$.

The eigenfunctions and eigenenergies of the effective Hamiltonian~\eref{hamil.eff} are calculated via 
\begin{equation}
\mathbf{U}^{-1} \mathcal{H}_{\mathrm{eff}} \mathbf{U} =\mathbf{W} \,.
\label{eq.diag}
\end{equation}
The columns of the unitary matrix $\mathbf{U}$ are the eigenvectors $|\psi_n\rangle (B)$ in the Hund's case~(b) basis, and $\mathbf{W}$ is a diagonal matrix consisting of the eigenvalues $E_n (B)$, where $n$ is a collective quantum number of the spin-rotation states.
The strength $B$ of the magnetic field enters as a parameter.
Equation~\eref{eq.diag} is solved numerically using built-in functions of the \textsc{Matlab} program package~\cite{matlab11b}.


\subsection{Dynamics in the optical centrifuge}

The optical centrifuge is created by two counter-rotating circularly polarized laser fields that are linearly chirped with respect to each other~\cite{karczmarek99,villeneuve00}.
The result is a linearly polarized field, where the polarization direction rotates with an increasing angular frequency.
The central frequency of the laser pulses is assumed to be far detuned from any molecular resonances, and the laser pulse interacts with the molecules via Raman-type excitations of the rotational levels~\cite{zon75,friedrich95,friedrich95b}.
The non-resonant laser pulse induces a dipole in the molecule via its electric polarizability and then interacts with the induced dipole.
Averaging over the fast oscillations of the laser field leads to an effective interaction potential~\cite{boyd08}
\begin{eqnarray}
V&=-\frac{\Delta \alpha}{4} E_0^2 \mathcal{E}^2(t) \left[ \bi{r} \cdot \bi{p}(t) \right]^2 \nonumber\\
&=-U_0 \mathcal{E}^2(t) \left[ \bi{r} \cdot \bi{p}(t) \right]^2 \,.
\label{eq.potential}
\end{eqnarray}
Here, $\Delta \alpha$ is the polarizability anisotropy, $E_0$ is the peak strength of the laser electric field, $\mathcal{E}(t)$ is the (dimensionless) envelope of the laser electric field, $\bi{r}=(r_x,r_y,r_z)$ is the orientation of the molecular axis, $\bi{p}$ is the (slowly rotating) polarization axis of the laser electric field, and $U_0=(\Delta \alpha E_0^2/4)$ is the trap depth.

The evolution of the wave function is calculated by solving the time-dependent Schr\"odinger equation,
\begin{equation}
i\hbar\frac{\partial |\Psi\rangle}{\partial t} = \left(\mathcal{H}_{\mathrm{eff}} +V \right) |\Psi\rangle \,,
\label{eq.tdse}
\end{equation}
using the effective spin-rotational Hamiltonian $\mathcal{H}_{\mathrm{eff}}$ [\eref{hamil.eff}-\eref{eq.spinzeeman}] and the laser interaction potential $V$ from~\eref{eq.potential}.
It is convenient to expand $|\Psi\rangle$ using the eigenfunctions $|\psi_n\rangle$ and eigenvalues $E_n$ of $\mathcal{H}_{\mathrm{eff}}$:
\begin{equation}
|\Psi(t)\rangle=\sum_{n} C_n(t) e^{-iE_nt/\hbar} |\psi_n\rangle \,.
\label{eq.expansion}
\end{equation}
We insert~\eref{eq.potential} and~\eref{eq.expansion} into~\eref{eq.tdse} to obtain a set of coupled differential equations for the expansion coefficients,
\begin{eqnarray}
\frac{\partial C_n (t)}{\partial t} = \frac{1}{i\hbar} \sum_m C_m(t) e^{-i(E_m-E_n)t/\hbar} \langle \psi_n | V(t) | \psi_m\rangle \,.\nonumber\\
\label{eq.differential}
\end{eqnarray}
For the temporal envelope of the electric field we use
\begin{eqnarray}
\mathcal{E}^2(t)=\left\{ \begin{array}{lll}
\sin^2\left[\pi\frac{t}{2 t_{\mathrm{on}}}\right] & \mathrm{for} & 0\le t < t_{\mathrm{on}} \\
1 & \mathrm{for} & t_{\mathrm{on}} \le t < {t_{\mathrm{p}}}-t_{\mathrm{off}} \\
\sin^2\left[\pi\frac{(t-t_{\mathrm{p}})}{2 t_{\mathrm{off}}}\right] & \mathrm{for} & {t_{\mathrm{p}}}-t_{\mathrm{off}} \le t < t_{\mathrm{p}} \\
0 & \mathrm{else.} &
\end{array}
\right.
\nonumber\\
\label{eq.envelope}
\end{eqnarray}
Here, $t_{\mathrm{on}}$ and $t_{\mathrm{off}}$ are the turn-on and turn-off time of the pulse, respectively, and $t_{\mathrm{p}}$ is the duration of the whole centrifuge pulse (including the turn-on and turn-off).
We consider a constant intensity of the field (apart from turn-on and turn-off), in order to gain a clear picture of the magneto-rotational effect, not blurred by molecules lost early from the centrifuge trap.
A realistic experimental envelope is generally slowly decaying~\cite{spanner01}.
The polarization vector of the electric field is rotating in the $yz$-plane, and is given as
\begin{equation}
\bi{p} (t)=\mathbf{\hat y} \sin\left(\beta\frac{t^2}{2}\right) + \mathbf{\hat z} \cos\left(\beta\frac{t^2}{2}\right)\,,
\label{eq.polarization}
\end{equation}
where $\mathbf{\hat y}$ and $\mathbf{\hat z}$ are unit vectors along the $y$ and $z$ axis, respectively, and $\beta$ is the acceleration of the centrifuge rotation.
Using~\eref{eq.potential}, \eref{eq.envelope}, and~\eref{eq.polarization}, the interaction potential can be expressed as
\begin{eqnarray}
V(t)=-U_0\mathcal{E}^2(t)\Bigg[r_y^2\sin^2\left(\beta\frac{t^2}{2}\right)
\nonumber\\
+ r_yr_z\sin\left(2\beta\frac{t^2}{2}\right) + r_z^2\cos^2\left(\beta\frac{t^2}{2}\right)\Bigg] \,.
\end{eqnarray}
We use the following relations to express $V(t)$ in terms of the rotation matrices $D_{mk}^{(j)*}$:
\begin{subequations}
\label{eq.xyzinD}
\begin{eqnarray}
r_x^2&=\frac{1}{3}-\frac{1}{3}D_{00}^{(2)*}+\frac{1}{\sqrt{6}}D_{20}^{(2)*}+\frac{1}{\sqrt{6}}D_{-20}^{(2)*} \\
r_y^2&=\frac{1}{3}-\frac{1}{3}D_{00}^{(2)*}-\frac{1}{\sqrt{6}}D_{20}^{(2)*}-\frac{1}{\sqrt{6}}D_{-20}^{(2)*} \\
r_z^2&=\frac{1}{3}+\frac{2}{3}D_{00}^{(2)*} \\
r_yr_z&=\frac{i}{\sqrt{6}}D_{10}^{(2)*}+\frac{i}{\sqrt{6}}D_{-10}^{(2)*}
\end{eqnarray}
\end{subequations}
The matrix elements of the rotation matrices in Hund's case~(b) basis are given as
\begin{eqnarray}
\langle JNM|&D_{m0}^{(2)*}|JNM \rangle=(-1)^{J+J'-M+1}
\nonumber\\
&\times\sqrt{(2J+1)(2J'+1)(2N+1)(2N'+1)}
\nonumber\\
&\times
\left(\begin{array}{ccc}J&2&J'\\-M&m&M'\end{array}\right)
\left(\begin{array}{ccc}N&2&N'\\0&0&0\end{array}\right)
\nonumber\\
&\times
\left\{\begin{array}{ccc}N'&J'&1\\J&N&2\end{array}\right\} \,.
\label{eq.Delements}
\end{eqnarray}
Mapping~\eref{eq.Delements} to the basis $|\psi_n\rangle$ by the use of $\bi{U}$ [see~\eref{eq.diag}] and using~\eref{eq.xyzinD}, we obtain the matrix elements $\langle \psi_n|V(t)|\psi_m \rangle$ of the laser induced potential.
Finally, the coupled differential equations~\eref{eq.differential} are solved numerically to obtain $|\Psi(t)\rangle$.
Note that the coefficients $C_n(t)$ are constant after the turn-off of the centrifuge pulse, so the wave function after the centrifuge pulse is
\begin{equation}
|\Psi(t>t_{\mathrm{p}})\rangle=\sum_{n} C_n(t_{\mathrm{p}}) e^{-iE_nt/\hbar} |\psi_n\rangle \,.
\end{equation}

To include thermal effects, we do ensemble averaging.
Therefore, we solve~\eref{eq.differential} for all thermally populated states $|\psi_n\rangle$, and average the result weighted by the respective Boltzmann factors.
To make the calculation feasible, we include only 95\% of the thermal population.
The density matrix is therefore given as
\begin{equation}
\varrho (t)=\mathcal{N}\sum_{i=1}^{i_{\mathrm{max}}} g_i |\Psi_i(t)\rangle\langle \Psi_i(t)| \,,
\label{eq.rho}
\end{equation}
where $\mathcal{N}$ is a normalization factor, $|\Psi_i(t)\rangle$ is the solution of~\eref{eq.tdse} for the initial state $|\Psi(0)\rangle=|\psi_i\rangle$, $g_i$ is the respective Boltzmann factor of the initial state, and $i_{\mathrm{max}}$ is chosen such that 95\% of the thermal population is included.


\subsection{Angular distribution}

In order to calculate the angular distribution $P(\theta,\phi)$ of the molecular axis ($\theta$ and $\phi$ being the polar and azimuthal angle, respectively), it is beneficial to use the Hund's case (a) basis.
In Hund's coupling case (a), the total angular momentum $\bi{J}$ is formed from its projection $\Omega$ on the molecular axis and from the orbital angular momentum $\bi{O}$ of the nuclei.
In particular, for an oxygen molecule in its electronic ground state, $\Omega$ is equal to the projection $\Sigma$ of the electronic spin on the molecular axis, where $\Sigma=0,\pm1$.
The eigenfunctions of Hund's case (a) are the symmetric top wave functions~\cite{vanvleck29}.
The internal axis of this symmetric top is the molecular axis, and the projection of $\bi{J}$ on the molecular axis is $\Sigma$.
The angular distribution $P(\theta,\phi)$ is therefore obtained from the wave function in Hund's case (a) by integrating over the internal coordinate, which yields
\begin{eqnarray}
P(\theta,\phi)=&\sum_{J,J',M,M',\Sigma} \varrho^{\mathrm{(a)}}_{JM\Sigma;J'M'\Sigma}(t)
\nonumber\\
&\times
\Theta^*_{J'M'\Sigma}(\theta)\Theta_{JM\Sigma}(\theta)
\Phi^*_{M'}(\phi)\Phi_{M}(\phi) \,,
\label{eq.angdist}
\end{eqnarray}
where $\varrho^{\mathrm{(a)}}_{JM\Sigma;J'M'\Sigma}$ are the elements of the density matrix in Hund's case~(a) basis and
\begin{subequations}
\begin{eqnarray}
\Theta_{JM\Sigma}(\theta)=&\sqrt{\frac{2J+1}{2}}d_{M\Sigma}^{(J)}(\theta)\\
\Phi_M(\phi)=&\frac{1}{\sqrt{2\pi}}e^{iM\phi} \,,
\end{eqnarray}
\end{subequations}
where $d_{M\Sigma}^{(J)}(\theta)$ are the reduced Wigner matrix elements.
They are given as~\cite{brown03}
\begin{eqnarray}
d_{M,\Sigma}^{(J)}(\theta)=
\nonumber\\
\sum_s (-1)^s \frac{\sqrt{(J+M)!(J-M)!(J+\Sigma)!(J-\Sigma)!}}{(J+M-s)!(J-\Sigma-s)!s!(s+\Sigma-M)!}
\nonumber\\
\times
\left(\cos\frac{\theta}{2}\right)^{2J+M-\Sigma-2s}\left(\sin\frac{\theta}{2}\right)^{2s-M+\Sigma}\,.
\label{eq.reducedWig}
\end{eqnarray}
The index $s$ runs over all values for which the arguments of the factorials are not negative.

Unfortunately, \eref{eq.reducedWig} is numerically hard to evaluate for most values of $M$ due to the large factorials involved.
Therefore, the following recursion relation is used:
\begin{subequations}
\label{eq.recursion}
\begin{eqnarray}
d_{M,\Sigma}^{(J)}(\theta)=&-2\frac{\Sigma-(M-1)\cos\theta}{\sin\theta\sqrt{(J-M+1)(J+M)}}d_{M-1,\Sigma}^{(J)}(\theta)
\nonumber\\
&-\sqrt{\frac{(J+M-1)(J-M+2)}{(J-M+1)(J+M)}}d_{M-2,\Sigma}^{(J)}(\theta)
\nonumber\\
\label{eq.recursionUp}\\
d_{M,\Sigma}^{(J)}(\theta)=&-2\frac{\Sigma-(M+1)\cos\theta}{\sin\theta\sqrt{(J+M+1)(J-M)}}d_{M+1,\Sigma}^{(J)}(\theta)
\nonumber\\
&-\sqrt{\frac{(J-M-1)(J+M+2)}{(J+M+1)(J-M)}}d_{M+2,\Sigma}^{(J)}(\theta) \,.
\nonumber\\
\label{eq.recursionDown}
\end{eqnarray}
\end{subequations}
This recursion relation is only stable for~\cite{prezeau10}
\begin{equation}
J(J+1)+\frac{2M\Sigma\cos\theta-M^2-\Sigma^2}{\sin^2\theta}\geq0 \,.
\label{eq.stability}
\end{equation}
When the inequality~\eref{eq.stability} is not fulfilled, the recursion has to be done in a direction which leads towards fulfilling the inequality~\cite{prezeau10}.
For $\Sigma=0$, this means using the recursion relation~\eref{eq.recursion} with decreasing $|M|$, in particular~\eref{eq.recursionUp} for negative $M$ and~\eref{eq.recursionDown} for positive $M$.
This scheme can also be used for the other two cases of $\Sigma=\pm1$ with an error negligible for our purposes.
The starting points for the recursion are [obtained from~\eref{eq.reducedWig}]
\begin{subequations}
\begin{eqnarray}
d_{J+1,\Sigma}^{(J)}(\theta)&=d_{-J-1,\Sigma}^{(J)}(\theta)=0 \\
d_{J,\Sigma}^{(J)}(\theta)&=(-1)^{J-\Sigma} \nonumber\\
&\times\sqrt{\left(\begin{array}{c}2J\\J+\Sigma\end{array}\right)}
\left(\cos\frac{\theta}{2}\right)^{J+\Sigma}\left(\sin\frac{\theta}{2}\right)^{J-\Sigma}
\nonumber\\
\\
d_{-J,\Sigma}^{(J)}(\theta)&=\sqrt{\left(\begin{array}{c}2J\\J+\Sigma\end{array}\right)}
\left(\cos\frac{\theta}{2}\right)^{J-\Sigma}\left(\sin\frac{\theta}{2}\right)^{J+\Sigma}
\,.
\nonumber\\
\end{eqnarray}
\end{subequations}
A test of the accuracy of the numerical method via the orthonormality of the obtained symmetric top eigenfunctions showed an error of less than 100~ppm, for angular momenta up to $J=42$.
This is sufficient for our purposes.

To obtain the angular distribution from~\eref{eq.angdist}, one needs to map the density matrix~\eref{eq.rho} to the basis of Hund's case~(a).
In particular, the mapping is given as~\cite{brown03}
\begin{eqnarray}
\langle J M \Sigma | J N M \rangle = (-1)^{J-1}\sqrt{2N+1}
\left(\begin{array}{ccc}J&1&N\\\Sigma&-\Sigma&0\end{array}\right) \,.
\nonumber\\
\end{eqnarray}


\subsection{\label{sec.raman}Raman signal}

The ultrafast rotation of the centrifuged molecules can be directly measured by Raman spectroscopy, using a circularly polarized probe pulse~\cite{korobenko13}.
The coherence between states with $\Delta N=\pm2$ causes a Raman shift of twice the rotational frequency of the centrifuged molecules.
The shift is negative (positive) if the molecules rotate in the same direction (opposite direction) as the probe pulse.
This frequency shift can be interpreted as a rotational Doppler effect~\cite{korech13}.

The probe signals are calculated using the model described in~\cite{steinitz14}.
The output $\bi{E}_{\mathrm{out}}$ of the probe field (with left and right circular polarized components) is given as
\begin{eqnarray}
\bi{E}_{\mathrm{out}}(d)\approx e^{i\omega_it} \exp(i\beta_i Gd)
\nonumber\\
\times
\left[\cos\left(\beta_i|\rho|d\right)\bi{I} + i\sin\left(\beta_i |\rho|d\right)\left(
\begin{array}{cc}
0&e^{-i\Phi}\\e^{i\Phi}&0
\end{array}
\right) \right]\bi{E}_{\mathrm{in}} \,,
\nonumber\\
\label{eq.raman}
\end{eqnarray}
where $\bi{E}_{\mathrm{in}}$ is the input of the probe field, $d$ is the length of the gas cell, $\beta_i$ is a parameter depending on the experimental conditions, $\omega_i$ is the central frequency of the probe, $\bi{I}$ is the identity matrix, $G$ is the expectation value of the alignment to the laser propagation axis, and $\rho=|\rho|e^{i\Phi}$ can be understood as a ``circular alignment'' in the centrifuge plane.
We use $d\beta_i=10$ in our calculations, as this approximately matches typical experimental conditions~\cite{korobenko13}.
For the coordinate system from figure~\ref{fig.centrifuge}, the quantities $G$ and $\rho$ are given as~\cite{steinitz14}
\begin{subequations}
\label{eq.Grho}
\begin{eqnarray}
G&=\langle r_x^2 \rangle \\
\rho&=\langle (r_y+ir_z)^2 \rangle \,.
\end{eqnarray}
\end{subequations}
The simulated probe signal is obtained from~\eref{eq.raman}.
The necessary expectation values are obtained as 
\begin{subequations}
\begin{eqnarray}
G(t)&= \mathrm{Tr}\left\{\mathbf{G}\varrho (t)\right\} \\
\rho(t)&=\mathrm{Tr}\left\{\brho\varrho (t) \right\} \,.
\end{eqnarray}
\end{subequations}
Here, $\varrho$ is the density matrix, and $\mathbf{G}$ and $\brho$ are the matrix representations of $G$ and $\rho$ [see~\eref{eq.Grho}].
The matrix elements of $\mathbf{G}$ and $\brho$ are obtained by the help of~\eref{eq.Grho}, \eref{eq.xyzinD} and~\eref{eq.Delements}.
As input probe we use a clockwise polarized pulse with a sine square envelope.
The measured signal is the anticlockwise part of the output of the probe.
It is given as a function of the pump-probe delay $t_{\mathrm{p}}$ as
\begin{equation}
S(t,t_{\mathrm{p}})= i e^{i\omega_it} e^{i10G} \sin\left(10 |\rho|\right) e^{-i\Phi} \mathcal{E}_{in}(t-t_{\mathrm{p}}) \,,
\label{eq.probe}
\end{equation}
where $\mathcal{E}_{\mathrm{in}}$ is the envelope of the probe pulse.
Finally, we take the Fourier transform over $t$ to obtain the output spectrum $|\tilde S(\omega,t_{\mathrm{p}})|^2$.



\section{\label{sec.results}Results of the simulations}

In this section, the results of the simulations are presented.
The parameters were chosen such that they closely resemble the experiment described in~\cite{milner14b}.
We consider a centrifuge pulse that is anticlockwise rotating around the $x$-axis, and therefore excites a rotational wavepacket with the orbital angular momentum $\bi{N}$ pointing along $-x$.
The duration of the laser pulses is $t_{\mathrm{p}}=36~\mathrm{ps}$, with an acceleration of $\beta=0.52\cdot10^{24}~\mathrm{s}^{-2}$ and a turn-on and turn-off time of $t_{\mathrm{on}}=t_{\mathrm{off}}=2~\mathrm{ps}$.
The intensity of the laser pulse is $I_0=2~\mathrm{TW}/\mathrm{cm}^2$, which corresponds to a trap depth of $U_0\approx16~B_0$ (where $B_0$ is the rotational constant).
At the end of the pulse, the centrifuge is spinning with 3~THz, which corresponds to a rotational level of $N\approx35$.
The initial rotational temperature of the molecules is set to room temperature, $T=298~\mathrm{K}$.
The strength of the magnetic field is $B=3~\mathrm{T}$, and its direction -- defining the $z$-axis -- is perpendicular to the laser-propagation axis.
The simulated probe pulses are right-circularly polarized, and have a duration of 10~ps (full width at half maximum), corresponding to a resolution of approximately 3~cm$^{-1}$.

\begin{figure}
\centering
\includegraphics[width=\linewidth]{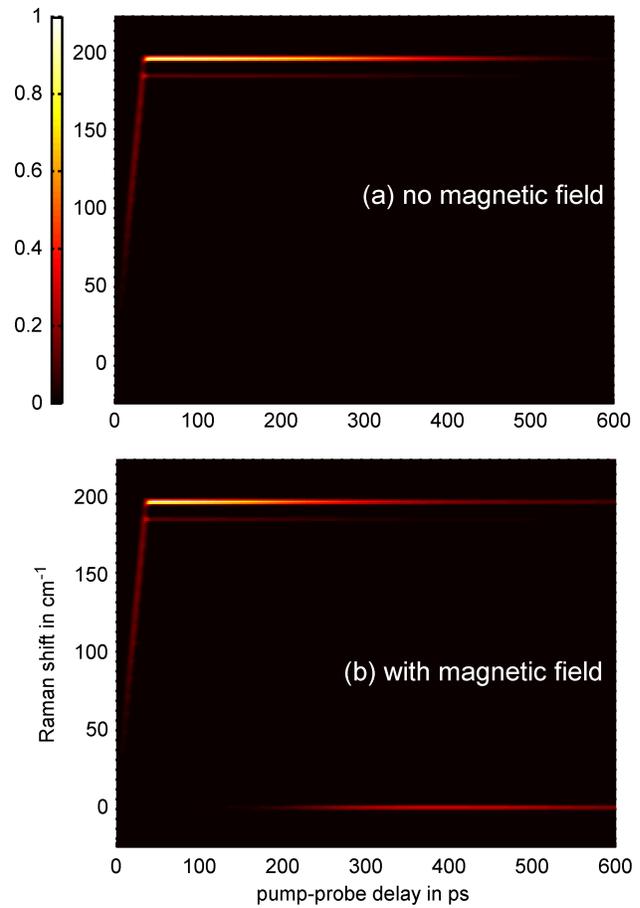}
\caption{
\label{fig.probe}
Probe spectrum for centrifuged oxygen molecules.
In panel~(a) the case of no magnetic field is shown, in panel~(b) the case for a magnetic field of 3~T.
The colour axis is in arbitrary units.
See text for the parameters.
}
\end{figure}

Figure~\ref{fig.probe} shows the simulated spectrum.
For comparison, also the case of no magnetic field is presented.
The simulation reproduces recent experimental results very well~\cite{korobenko13,milner14b,milner14c}.
During the centrifuge pulse (the first 36~ps), whilst the molecules are accelerated, one can see a strong anti-Stokes signal (the centrifuge and the probe pulse are rotating in opposite directions) with a linearly increasing frequency shift.
After the end of the laser pulse, the signal remains at a constant shift and consists of two lines around 200 cm$^{-1}$.
This shift is caused by molecules rotating anticlockwise around the $x$-axis.
In particular, the two lines correspond to the levels $N=35$ and $N=33$ (remember that the shift is twice the rotational frequency, see section~\ref{sec.raman}).
After about 500~ps, the Raman signal fades.
This is due to the coherent dephasing of the three spin states~\cite{milner14c}.
For the case of no magnetic field, the Raman signal partially revives after about 1~ns, and a full revival can be observed after about 2~ns (not shown in figure~\ref{fig.probe}).
If a magnetic field is applied, no such revivals are observed.
However, another signal appears:
After about 300~ps, a Rayleigh component becomes visible.
Such a line has also been observed in a recent experiment~\cite{milner14b}.

\begin{figure}
\centering
\includegraphics[width=\linewidth]{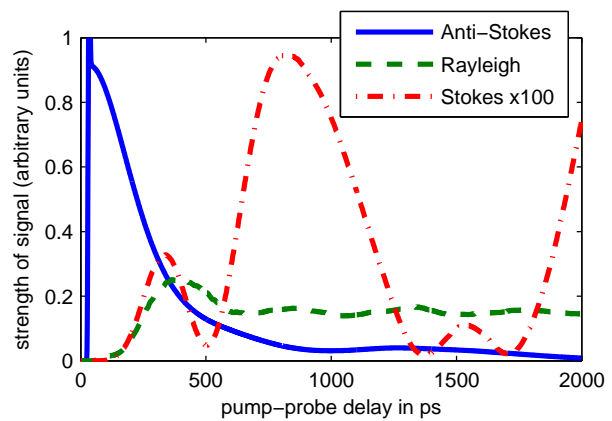}
\caption{
\label{fig.integrated}
Relative strengths of the probe signal components (anti-Stokes, Rayleigh, and Stokes) for centrifuged molecules in a magnetic field of 3T.
The strong peak of the anti-Stokes component during the centrifuge pulse (the first 36 ps) is due to the strong alignment in the centrifuge field.
See text for the parameters.
}
\end{figure}

Figure~\ref{fig.integrated} shows the strength of the three parts (Stokes, Rayleigh, and anti-Stokes) of the probe signal, for pump-probe delays of up to 2~ns.
It can be seen that the anti-Stokes signal decreases almost monotonically, with a weak local maximum at around 1300 ps.
The Rayleigh signal slowly builds up, reaching a maximum after 400~ps, then remaining approximately constant.
Remarkably, also a Stokes signal builds up over time.
It reaches a maximum at 800~ps and is about hundred times weaker than the other two signals.
The appearance of the Stokes signal indicates that some of the molecules may have changed their direction of rotation.
An experimental observation of this Stokes signal has also been reported in~\cite{milner14b}.
For the case of no magnetic field (not shown here), neither a Stokes nor a Rayleigh signal can be observed.

\begin{figure*}
\includegraphics[width=\linewidth]{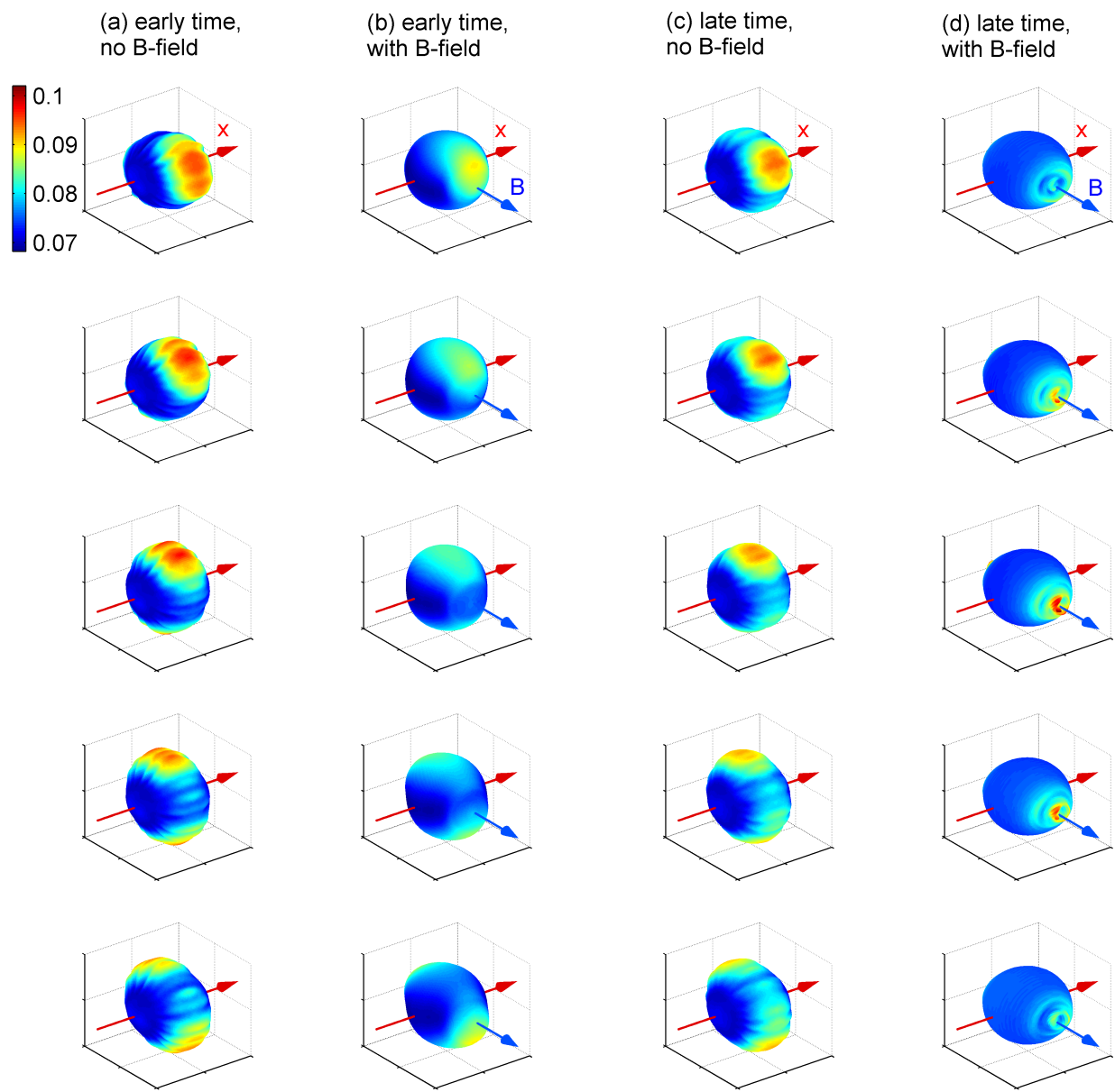}
\caption{
\label{fig.angulardist}
Angular distribution of the centrifuged oxygen molecules at different times, (a,b) without magnetic field and (c,d) with a field of $B=3~\mathrm{T}$.
The red arrow depicts the propagation axis of the centrifuge pulse, the blue arrow the magnetic field.
Time evolves from top to bottom rows, with time steps of 25~fs per row.
The first row is at $t=36~\mathrm{ps}$ (right after the end of the centrifuge laser pulse) for early times (a,c), and at $t=1000~\mathrm{ps}$ for late times (b,d).
See text for the parameters.
See also the movie in the supplementary material.
}
\end{figure*}

Additional insights into the dynamics can be gained from the angular distribution.
It is shown in figure~\ref{fig.angulardist} for different times after the end of the centrifuge laser pulse (a short movie can be found in the supplementary material).
For no magnetic field, we can see that the distribution is elongated along an axis that rotates anticlockwise around the laser propagation axis.
The dynamics does not change significantly over time [compare figures~\ref{fig.angulardist}~(a) and~(b)].
The picture is quite different when a magnetic field is applied.
Shortly after the centrifuge pulse [figure~\ref{fig.angulardist}~(c)], we can see approximately the same distribution as for $B=0$.
At later times [figure~\ref{fig.angulardist}~(d)], a very different picture emerges.
Whilst the angular distribution is still elongated, it now points along the magnetic field and is not rotating any more.
This alignment along the $z$-axis can explain the appearance of the Rayleigh component in the probe signal.



\section{\label{sec.mechanism}Rotation of the angular momentum by the magnetic field}

\begin{figure*}
\includegraphics[width=\linewidth]{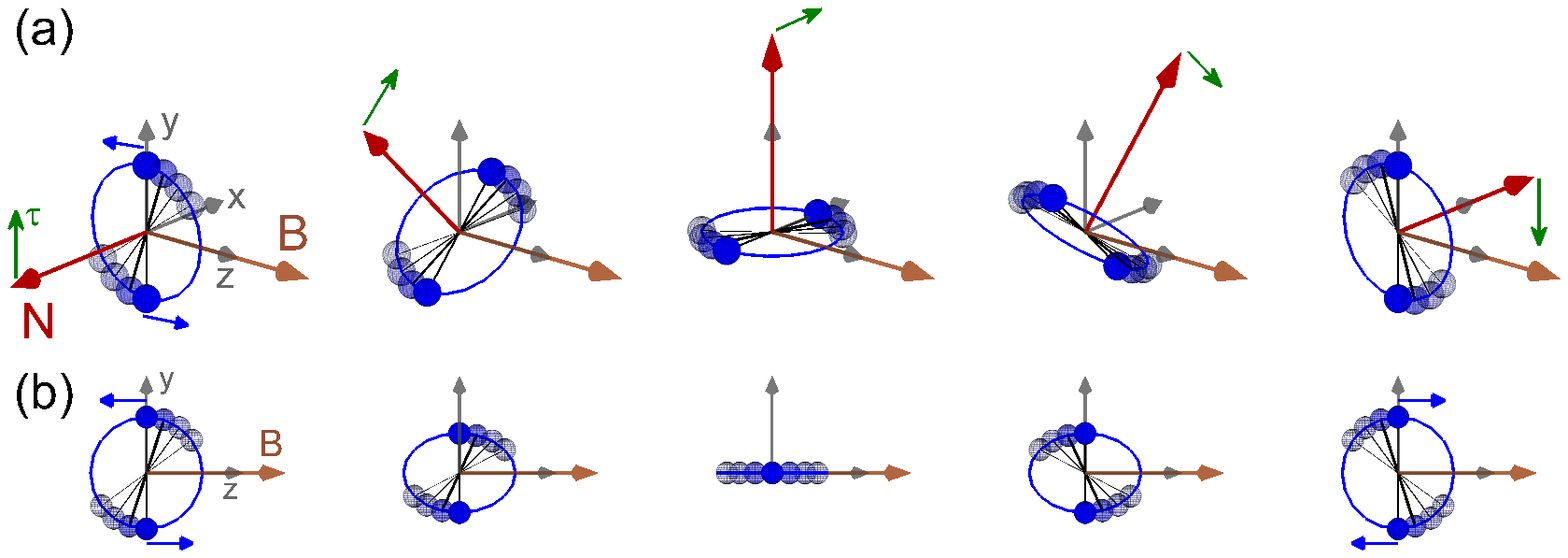}
\caption{
\label{fig.rotationPic}
(a) A sketch of the proposed mechanism of the magneto-rotational effect.
The panels show the situation at different times (time progressing from left to right).
Shortly after the end of the centrifuge laser pulse, the orbital angular momentum $\bi{N}$ points along $-x$, and the molecules rotate anticlockwise in the $yz$-plane.
A magnetic field induced torque $\btau$ causes a precession of $\bi{N}$ around the magnetic field $\bi{B}$.
The plane of rotation of the molecules (indicated by the blue ring) precesses accordingly.
In panel (b), the projections of the pictures of panel (a) on the $yz$-plane (the polarization plane of the probe pulse) are shown.
It can be seen that at different times the molecules rotate clockwise or anticlockwise, or they appear to be aligned along the magnetic field.
}
\end{figure*}

In order to explain the observed birefringence and Stokes signals, it was proposed in~\cite{milner14b} that the orbital angular momentum~$\bi{N}$ precesses around the magnetic field $\bi{B}$.
This scenario is depicted in figure~\ref{fig.rotationPic}~(a).
Right after the centrifuge pulse (leftmost panel), $\bi{N}$ points along $-x$ and the molecules rotate in the $yz$-plane.
A magnetic field induced torque $\btau$ acts on $\bi{N}$, leading to a precession of $\bi{N}$ around $\bi{B}$.
The situation as seen for the probe pulse is shown in figure~\ref{fig.rotationPic}~(b), where the projection onto the $yz$-plane is shown.
At the beginning (leftmost panel) a anticlockwise rotation of the molecules in the $yz$-plane is seen.
Accordingly, the probe signal has a strong anti-Stokes component (remember that the probe is right-circularly polarized).
After some time, when $\bi{N}$ points along the $y$-axis, the molecules rotate in the $xz$-plane.
Projected onto the $yz$-plane, this is seen as alignment along the $z$-axis (middle panel).
Accordingly, the probe signal acquires a Rayleigh component.
Eventually, $\bi{N}$ points along $+x$ and uni-directional rotation is seen again in the $yz$-plane -- but now in a clockwise sense (rightmost panel).
This leads to the observation of a Stokes component in the probe spectrum.

The torque $\btau$ can not originate from a direct Zeeeman interaction between $\bi{B}$ and $\bi{N}$.
The respective g-factor is of the order of $10^{-4}$~\cite{christensen78}; for the considered field strengths of a few Tesla, this would correspond to precession periods of 100~ns and longer, clearly too slow to explain the observed effect (such a slow precession has been observed experimentally for nitrogen superrotors~\cite{korobenko15a}).
Instead, the interaction between $\bi{N}$ and $\bi{B}$ is mediated by the electronic spin $\bi{S}$.
The spin interacts with the magnetic field via the Zeeman interaction $\mathcal{H}_{\mathrm{ms}}$, and with $\bi{N}$ via the spin-rotation interactions $\mathcal{H}_{\mathrm{ss}}$ and $\mathcal{H}_{\mathrm{so}}$.
As a result, the effective Hamiltonian depends on the orientation of $\bi{N}$ with respect to $\bi{B}$, giving rise to the torque $\btau$.

In order to test the proposed model, we calculate the frequency of the precession of $\bi{N}$ around $\bi{B}$.
The torque $\btau$ acting on $\bi{N}$ is given as
\begin{equation}
\btau=-\bi{N} \times \left(\bnabla_{\bi{N}} \cdot W_{N,\sigma}(B) \right) \,,
\label{eq.torque}
\end{equation}
where $W_{N,\sigma}(B)$ is the potential energy arising from the magnetic field, for a molecule in the rotational level $N$ and spin-state $\sigma$ (We introduced the new label $\sigma$ for the spin-state, as the meaning of $\sigma$ changes with the strength of the magnetic field).
Since the magnetic field points along the $z$-axis, we obtain from~\eref{eq.torque} the precession frequency $\nu_{N,\sigma}$ as
\begin{equation}
\nu_{N,\sigma}=\frac{1}{h}\frac{\partial W_{N,\sigma}(B)}{\partial N_z} \,,
\label{eq.nu}
\end{equation}
where $h$ is Planck's constant, and $N_z$ is the projection of $\bi{N}$ on the $z$-axis.
In the following, we will derive analytical expressions of $\nu_{N,\sigma}$ as a function of the field strength and the terminal velocity of the centrifuge, for two limiting cases:
The case of a weak magnetic field, i.e. when the spin-Zeeman interaction is much weaker than the spin-rotation interaction, and the opposite case of a strong magnetic field.

\subsection{Weak magnetic field}

For a weak magnetic field, the electronic spin is coupled strongly to the orbital angular momentum; the magnetic field acts as a perturbation to the precession of $\bi{S}$ around $\bi{N}$.
The spin-rotation interaction is therefore independent of the orientation of $\bi{N}$ and the only non-zero term in $\nu_{N,\sigma}$ arises from the spin-Zeeman term,
\begin{equation}
\nu_{N,\sigma} \approx \frac{g_S \mu_B}{h}\frac{\partial \left(\bi{B} \cdot \bi{S}\right)}{\partial N_z}  = \frac{g_S \mu_B}{h} B \frac{\partial S_z}{\partial N_z} \,.
\end{equation}
Here, $S_z$ is the projection of $\bi{S}$ on the space-fixed $z$-axis.
Averaging over the fast rotation of $\bi{S}$ around $\bi{N}$, we obtain $S_z=S_NN_z/|\bi{N}|$, where $S_N\approx0,\pm1$ is the projection of $\bi{S}$ on $\bi{N}$ (note that $S_N$ characterizes the three spin states, thus $\sigma\equiv S_N$).
Thus, the precession frequency of $\bi{N}$ around the magnetic field is given as
\begin{equation}
\nu_{N,\sigma} \approx g_S \mu_B B \frac{S_N}{h |\bi{N}|} \,.
\label{eq.weak}
\end{equation}

Equation~\eref{eq.weak} shows that the precession frequency is proportional to the ratio $B/N$.
This stems from the fact that the torque, induced by the spin-Zeeman interaction, is proportional to the strength of the magnetic field, but independent of $N$; yet, a higher angular momentum needs a stronger torque to precess with the same frequency.
It can furthermore be seen that the torque is proportional to $S_N$, thus the three spin-states precess differently:
The $J=N\pm1$ states precess with the same period, but in opposite directions, and the $J=N$ state does not precess at all.
Also, $\nu_{N,\sigma}$ is independent of $N_z$, thus the distribution of the angular momentum for a single spin state does not disperse, but remains in a narrow cone that was created by the centrifuge.

For the angular distribution, averaged over the fast rotation of the molecular axis, we can conclude the following:
Right after the centrifuge pulse, it has the shape of a disc in the $yz$-plane.
In the time-scale of hundreds of picoseconds, this disc splits into three discs, corresponding to the three spin-states:
One remaining in the $yz$-plane, and two rotating in opposite directions around $z$.
In a very recent experiment, Milner~\textit{et al.}~\cite{korobenko15a} succeeded to observe this behaviour.

\begin{figure}
\centering
\includegraphics[width=\linewidth]{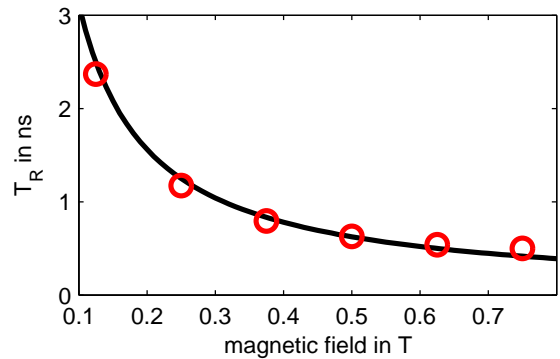}
\caption{
\label{fig.weak}
The time $T_{\mathrm{R}}$ of the first peak of the Rayleigh signal as a function of the strength $B$ of the magnetic field, for weak field strengths.
The solid line shows the predictions by~\eref{eq.weak} (with $N=35$), the circles are the results of simulations using the same parameters (apart from $B$) as in section~\ref{sec.results}.
}
\end{figure}

Since there is no dispersion, the Rayleigh signal should show clear beating with a frequency of $2\nu_{N,\sigma}$ (only the alignment, not the orientation of $\bi{N}$ is important for the signal).
The signal should first peak at $T_{\mathrm{R}}=1/(4\nu_{N,\sigma})$ after the end of the centrifuge pulse, when $\bi{N}$ points along the $y$-axis.
We compare $T_{\mathrm{R}}$ as predicted by the analytical formula~\eref{eq.weak} to numerical simulations via the Schr\"odinger equation~\eref{eq.tdse}.
This comparison is shown in figure~\ref{fig.weak}, where the solid line is the analytical prediction and the circles are the results of the simulations.
A very good agreement can be seen.
The simulations also confirmed the beating of the Rayleigh signal (not shown here).

If the model is correct, we should expect a similar behaviour for the Stokes signal, with beating at $\nu_{N,\sigma}$ and a first peak at $T_{\mathrm{S}}=1/(2\nu_{N,\sigma})$.
However, simulations revealed that this is not the case.
Furthermore, the signal is two orders of magnitude smaller than the anti-Stokes signal.
The likely reason is that the Stokes signal is -- differently to the Rayleigh signal -- affected by coherent dephasing of the different spin states.
The Stokes signal might also be much more sensitive to weak perturbations by the magnetic field that are not covered by the approximations we made in deriving~\eref{eq.weak}.

\subsection{Strong magnetic field}

For a strong magnetic field, the spin is strongly coupled to the field and precesses around it.
The three spin-states are defined by the projection of $\bi{S}$ on the magnetic field axis, $S_z=0,\pm1\equiv\sigma$.
The orbital angular momentum is indirectly affected by the magnetic field via the spin-orbit coupling $\mathcal{H}_{\mathrm{so}}$ and the spin-spin coupling $\mathcal{H}_{\mathrm{ss}}$.
The latter includes the orbital angular momentum implicitly via the projection $S_{\zeta}$ of the spin on the molecular axis.
Averaging over the fast rotation of the molecular axis yields $S_{\zeta}\approx 0.5 \sin^2\alpha$, where $\alpha$ is the angle between $\bi{S}$ and the nuclear orbital angular momentum $\bi{O}$.
Approximating $\bi{O}$ by $\bi{N}$~\cite{tinkham55a}, we can express the spin-spin interaction as
\begin{eqnarray}
\mathcal{H}_{\mathrm{ss}}&=\frac{2}{3}\lambda\left[3S_{\zeta}^2-\bi{S}^2\right]\nonumber\\
&\approx \lambda\left[1-\cos^2\alpha-\frac{2}{3}\bi{S}^2\right] \nonumber\\
&\approx \lambda\left[1-\frac{(\bi{S}\cdot\bi{N})^2}{N(N+1)}-\frac{2}{3}\bi{S}^2\right] \,.
\label{eq.ssapprox}
\end{eqnarray}
Adding the spin-orbit coupling, we obtain the spin-rotation coupling between $\bi{S}$ and $\bi{N}$ as
\begin{equation}
\mathcal{H}_{sr}\approx\gamma \bi{N}\cdot\bi{S}+\lambda\left[1-\frac{(\bi{S}\cdot\bi{N})^2}{N(N+1)}-\frac{2}{3}\bi{S}^2\right] \,.
\label{eq.sr}
\end{equation}
Averaging over the precession of $\bi{S}$ around the magnetic field $\bi{B}$ (which is much faster than the precession of $\bi{N}$ around $\bi{B}$) yields
\begin{eqnarray}
\mathcal{H}_{sr}\approx&\gamma N_zS_z-\lambda\frac{N_z^2}{N(N+1)}\left(\frac{3}{2}S_z^2-1\right)
\nonumber\\
&-\lambda\left(\frac{2}{3}-\frac{1}{2}S_z^2\right) \,.
\end{eqnarray}
Since $S_z$ and therefore the spin-Zeeman term is constant, \eref{eq.nu} becomes
\begin{equation}
\nu_{N,\sigma}\approx\frac{1}{h}\frac{\partial\mathcal{H}_{\mathrm{sr}}}{\partial N_z}=\frac{\gamma}{h}S_z-2\frac{\lambda}{h}\frac{N_z}{N(N+1)}\left(\frac{3}{2}S_z^2-1\right) \,.
\label{eq.strong}
\end{equation}

The precession frequency of $\bi{N}$ for the limit of a strong magnetic field includes two very different terms.
Both terms are of the same order of magnitude for typical oxygen superrotors ($30\lesssim N\lesssim100$).
The first one, stemming from the spin-orbit coupling, looks similar to the weak-field case.
It is again proportional to the approximate quantum number defining the spin state.
There is no $1/N$ dependence, since the spin-orbit coupling is proportional to the size of $\bi{N}$, and thus the torque increases with $N$.
The second term in~\eref{eq.strong} stems from the spin-spin interaction.
The latter depends quadratically on $N_z$, and thus the interaction does not average to zero for the $S_z=0$ state.
Therefore, in the strong field case all three spin states show a precession of $\bi{N}$.
Additionally, the dependence of $\nu_{N,\sigma}$ on $N_z$ means that the distribution of $\bi{N}$ disperses during the precession, leading to a more or less equal distribution of $\bi{N}$ in the $xy$-plane over time.
There is no dependence on the strength of the magnetic field in~\eref{eq.strong}.
This is not surprising, since the torque arises from the spin-rotation coupling;
this coupling is caused by intramolecular interactions and therefore independent of any external field.
An experimental observation of the independence of the magneto-rotational effect from $B$ for strong fields was reported in~\cite{milner14b}:
There, the birefringence signal was measured 1~ns after the centrifuge pulse as a function of the magnetic field strength.
It was found that the signal strength becomes independent of the magnetic field strength for sufficiently strong fields.

\begin{figure}
\centering
\includegraphics[width=\linewidth]{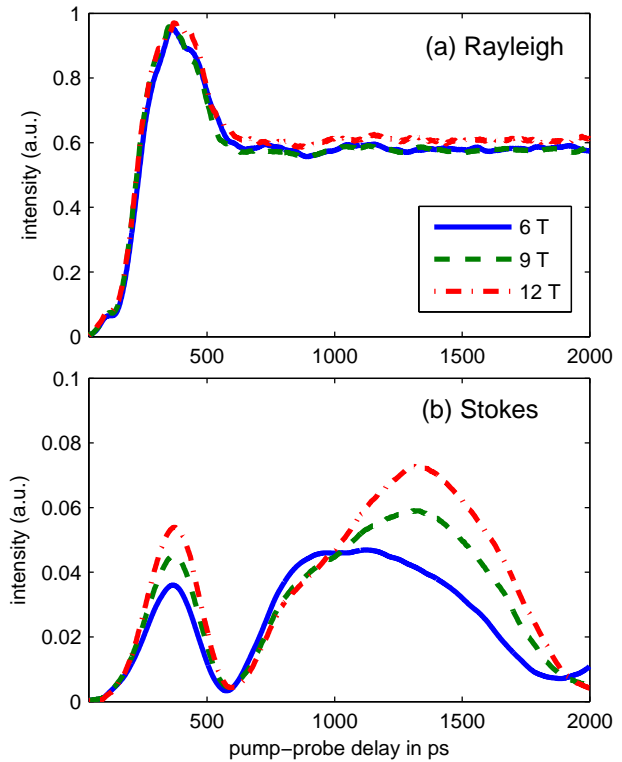}
\caption{
\label{fig.strong}
Shown are the relative strengths of the (a) Rayleigh and (b) Stokes signal for centrifuged molecules, for the case of strong magnetic fields (in particular, $B=6,9,12~\mathrm{T}$).
Note that the three shown Rayleigh signals are almost identical.
The parameters (apart from $B$) are the same as in section~\ref{sec.results}.
}
\end{figure}

We again test the predictions of~\eref{eq.strong} by comparing them to simulations.
The results for the Rayleigh signal are shown in figure~\ref{fig.strong}~(a), for the same parameters as in section~\ref{sec.results}, except for a stronger magnetic field.
Changing the field strength has almost no influence on the signal.
It furthermore remains more or less constant 500~ps after the centrifuge pulse, which is in accordance with $\bi{N}$ being equally distributed in the $xy$-plane.
There is a strong peak at 400~ps which we can not explain with the model.
The Stokes signal, shown in figure~\ref{fig.strong}~(b), is also approximately independent of $B$.



\section{Discussion and Conclusion}

In~\cite{milner14b} it was reported that paramagnetic superrotors interacting with a magnetic field $\bi{B}$ become aligned to the field.
As a mechanism, spin-mediated precession of the orbital angular momentum $\bi{N}$ (which is mainly the angular momentum of the nuclei) around the magnetic field was proposed.
In particular, the electronic spin $\bi{S}$ interacts with the magnetic field $\bi{B}$ via the spin-Zeeman interaction, and with the orbital angular momentum $\bi{N}$ via the spin-rotation coupling.
Thus, a strong torque stemming from the magnetic field acts on the orbital angular momentum.
This torque is orders of magnitude larger than the torque stemming from a direct Zeeman interaction between $\bi{N}$ and $\bi{B}$.

In this article, we conducted a thorough theoretical analysis of the magneto-rotational effect.
Starting from the proposed model, we derived analytical expressions for the precession frequency in the limits of weak and strong magnetic fields, respectively.
This allowed us also to make predictions about the distribution of $\bi{N}$ and the molecular axis.
For weak fields, a splitting of the angular momentum distribution into three parts (one for each spin state) was predicted, and the precession frequency is proportional to the strength of the magnetic field.
For strong fields, it was shown that the angular momentum becomes evenly distributed in the plane perpendicular to the magnetic field, and the precession is independent of the field strength.
A comparison of the analytical derivations on one hand, and numerical simulations (reported here) as well as experimental observations (reported in~\cite{milner14b,korobenko15a}) on the other hand, showed a quantitative agreement for the weak field case and a qualitative agreement for the strong field case.

The control of molecular rotation by electromagnetic fields is used in a large number of applications~\cite{lemeshko13}.
The magneto-rotational effect of paramagnetic superrotors adds an additional control knob for shaping the angular momentum distribution of molecules.
In particular for weak fields, a high degree of control is possible, including selective control of different spin states.
The analysis of the magneto-rotational effect provided in this article allows for proper tailoring of magnetic and centrifuge fields in possible applications.
The magneto-rotational effect shows qualitative changes for different field strengths; this makes the control of molecular trajectories by inhomogeneous magnetic fields~\cite{purcell09,gershnabel10,gershnabel11} a particularly interesting application for the magneto-rotational effect.

\ack

In the Averbukh's Light-Matter Interactions Group at the Weizmann Institute of Science, we had the privilege of enjoying scientific contacts with Moshe Shapiro over many years.
A great number of inspiring and insightful discussions with Moshe Shapiro on various aspects of the laser control of molecular dynamics have benefited directly and indirectly our work, including the one presented in this article.

I appreciate many fruitful discussions related to the problem of paramagnetic superrotors with Ilya Sh.~Averbukh, Aleksey Korobenko, Alexander A.~Milner, and Valery Milner.
A part of the research presented in this article was done during a visit to the University of British Columbia; I am very thankful to the group of Valery Milner for their kind hospitality, and to the UBC VPRI Mobility Award for the support.
I also thank Ilya Sh. Averbukh, Matthias Berg, and Axel Schild for a critical reading of the manuscript.
Financial support by the ISF (Grant No.~601/10), the DFG (Project No.~LE 2138/2-1), and the Minerva Foundation is gratefully acknowledged.
This research was made possible in part by the historic generosity of the Harold Perlman Family.



\section*{References}

\providecommand{\newblock}{}

\end{document}